\documentclass[12pt]{article}

\usepackage{amsmath}
\usepackage{amssymb}
\usepackage{amsbsy}
\usepackage{ifthen}
\usepackage{color}
\usepackage{graphicx}
\usepackage{float}
\usepackage{enumitem}
\usepackage{hyperref}
\usepackage{natbib}
\usepackage{setspace}
\usepackage{amsthm}
\usepackage{fullpage}
\usepackage{bbm}
\usepackage{multirow}
\usepackage{rotating}

\usepackage{float}
\usepackage{morefloats}
\usepackage[nolists]{endfloat}

\newcommand{\tr}{^{\prime}}
\def\b#1{\mbox{\boldmath $#1$}}    

\def\cg#1{\mbox{${\cal #1}$}}      
\def\cgl#1{\mbox{\scriptsize {${\cal #1}$}}}

\newcommand{\pa}{\partial}         
\renewcommand{\th}{\theta}

\newcommand{\al}{\alpha}
\newcommand{\be}{\beta}
\newcommand{\de}{\delta}
\newcommand{\la}{\lambda}

\newcommand{\ga}{\gamma}

\newcommand{\ba}{\begin{eqnarray}}
\newcommand{\ea}{\end{eqnarray}}

\usepackage{bm}
\usepackage{epsfig}
\usepackage{subfig}

\begin{document}

\title{\vspace*{-1cm} Marginal models with individual-specific effects for the analysis of longitudinal bipartite networks}
\author{Francesco Bartolucci\footnote{Department of Economics,  University of Perugia, Via A. Pascoli, 20, 06123 Perugia, Italy.} 
\footnote{{\em email}: francesco.bartolucci@unipg.it}
\and
Antonietta Mira\footnote{InterDisciplinary Institute of Data Science, Universit\`{a} della Svizzera Italiana, Switzerland, and
Department of Sciences and High Technology, Universit\`{a} degli Studi dell'Insubria, Italy} \footnote{{\em email}: antonietta.mira@usi.ch}
\and
Stefano Peluso\footnote{Department of Statistical Sciences,  Universit\`{a} Cattolica del Sacro Cuore, Italy.} \footnote{ {\em email}: stefano.peluso@unicatt.it}
}  

\maketitle
\begin{abstract}
\singlespacing
\noindent 
A new modeling framework for bipartite social networks arising from a sequence of partially time-ordered relational events is proposed.
We directly model the joint distribution of the binary variables indicating if each single actor is involved or not in an event.
The adopted parametrization is based on first- and second-order effects, formulated as in marginal models for categorical data and free higher order effects.
In particular, second-order effects are log-odds ratios with meaningful interpretation from the social perspective in terms of tendency to cooperate, in contrast to first-order effects interpreted in terms of tendency of each single actor to participate in an event.
These effects are parametrized on the basis of the event times, so that suitable latent trajectories of individual behaviors may be represented.
Inference is based on a composite likelihood function, maximized by an algorithm with numerical complexity proportional to the square of the number of units in the network.
A classification composite likelihood is 
used to cluster the actors, simplifying the interpretation of the data structure.
The proposed approach is illustrated on a dataset of scientific articles published in four top statistical journals from 2003 to 2012.
\vspace*{3mm}

\end{abstract}\newpage
\section{Introduction}\label{sec:intro}
In many relational contexts, a set of events is observed, with each event involving an arbitrary number of actors and even a single actor.
These events give rise to a so-called affiliation network in which there are two types of node: actors and events.
Following the current literature, see \cite{wang2009exponential} among others, we refer to this structure as {\em bipartite network}, also known as \textit{two-mode network}, contrarily to the \textit{one-mode network} having a unique type of nodes. 
An example, which motivates the present paper, is that of academic articles in top statistical journals \citep{ji2017coauthorship}, typically involving more than two authors but that might also be written by a single researcher.
In these applications, the interest is in studying the relations between units, with the aim of modeling separately the tendency of each unit to be involved in an event, and the tendency of each pair of units to cooperate.
We are also interested in studying the time evolution of social behaviors and thus the dynamics over time of both these tendencies.
The dataset of statistical publications we aim to analyze has also a particular feature that is important for the following developments: the events are only partially ordered as we know the year of publication of each article, but there is not any sensible way to temporally order articles published in the same year.

The literature on bipartite networks is mainly based on models having characteristics similar to those for one-mode networks in which direct connections are observed between certain pairs of actors, such as Exponential Random Graph Models \citep[ERGMs;][]{frank1986markov,wasserman1996logit}; for a review see \cite{snijders2011statistical} and \cite{amati2018social}.
One of the first model for the analysis of bipartite networks is proposed in \cite{iacobucci1990social} and is based on an ERGM structure with specific effects for both types of node (i.e., actors and events) and strong assumptions of independence between the response variables.
This approach was extended in several directions by \cite{skvoretz1999logit}, whereas \cite{wang2009exponential} presented a flexible class of ERMGs for bipartite networks and related estimation methods.
More recently, a review of models for this type of networks has been illustrated by \cite{aitkin2014statistical}, including certain versions of the  \cite{rasch:67} model and the latent class model \citep{good:74}.
The approach proposed in the present paper is also related to models for the analysis of longitudinal one-mode networks, such as actor-oriented models \citep{snijders1997simulation,snijders2010introduction}, dynamic ERGMs \citep{robins2001random}, hidden Markov models \citep{yang2011detecting,matias2015statistical,bartolucci2018dealing}, and the models for relational events described in \cite{dubois2013hierarchical}, \cite{perry2013point}, \cite{butts2017relational}, \cite{stadtfeld2017dynamic}, \cite{fox2016modeling}, and \cite{xia2016measuring}.

For the analysis of bipartite networks, and in particular of the dataset of publications is top statistical journals \citep{ji2017coauthorship}, we represent each event by a vector of response variables $\b Z^{(e)}=(Z_1^{(e)},\ldots,Z_n^{(e)})'$, with $Z_i^{(e)}$ equal to 1 if unit $i$ is involved in event $e$ and 0 otherwise.
Our aim is to directly formulate a statistical model for the response vectors $\b Z^{(e)}$ having a meaningful interpretation.
In particular, we rely on a marginal model \citep{bergsma2002marginal,bergsma2009marginal} based on first- and second-order effects.
The first-order effects correspond to the logit of the marginal distribution of each $Z_i^{(e)}$ variable and represent the general tendency of actor $i$ to be involved in event $e$.
The second-order effects are the log-odds ratios \citep[][Ch. 2]{agresti:2013} for the marginal distribution of each pair of variables $(Z_i^{(e)},Z_j^{(e)})'$, representing the tendency of actors $i$ and $j$ to be jointly involved in the same event $e$.
However, as we show in detail in the sequel, this parameter may be directly interpreted as the tendency of $i$ and $j$ to cooperate.
Moreover, even if we do not directly consider higher order effects, we do not pose any restrictions on these effects.
At least to our knowledge, the use of marginal models for the analysis of social network data is new in the statistical literature.

Second, we pay particular attention to the parametrization of the above effects so as to account for the time evolution, and represent individual trajectories in terms of tendency to participate in an event and tendency to cooperate.
This feature is common in latent growth models \citep{bollen2006latent}; however, in the proposed approach we use individual fixed parameters, rather than random parameters, applied to polynomials of time of suitable order.
Then, the proposed approach is particularly appropriate when the interest is in the evaluation of the behavior of a single actor in terms of the tendencies mentioned above. 
The possibility to estimate fixed parameters is possible thanks to the amount of information that is typically huge in the applications of interest.
For instance, in the motivating example, there are more than three thousand papers that play the role of events.

Third, in order to estimate the fixed individual parameters, we rely on a composite likelihood approach \citep{lindsay1988composite,varin2011overview}.
In particular, we use a likelihood function based on the marginal distribution of every ordered pair of actors.
For each of these pairs, the likelihood component directly depends on the first- and second-order effects described above, and on individual parameters referred to the two actors.
Then, to maximize the target function, we propose a simple iterative algorithm with $O(n^2)$ complexity, 
that is thus computationally tractable even with if the number of actors is large.
This is an important feature given the large scale of nowadays social network data; see also the discussion in \cite{vu2013model}.

Forth, in presence of many statistical units, we show how to  cluster units in groups that are homogenous in terms of tendency to be involved in an event or tendency to cooperate with other units.
For this aim, we rely on a classification composite likelihood function that is related to that used for estimating the individual fixed parameters.
This allows us to represent trajectories referred to homogeneous groups, rather than to individuals, so as to simplify the interpretation of the evolution of the data structure and of the social perspective of the phenomenon under study.

The paper is organized as follows.
In the next section we describe assumptions and interpretation of the proposed approach.
In Section \ref{sec:inference} we outline the method of inference based on the use of fixed effects and clustering techniques.
The application is illustrated in Section \ref{sec:app}.
In the last section we draw main conclusions and outline some possible extensions, as the inclusion of third-order effects and of individual covariates.

The estimation algorithm is implemented in a series of {\tt R} functions that we make available to the reader upon request.
\section{Proposed model}
Let $n$ denote the number of actors and $r$ the number of observed relational events.
Also let $Z_i^{(e)}$ be a binary outcome equal to 1 if the relational event $e$ involves unit $i$ and to 0 otherwise, with $i=1,\ldots,n$ and $e=1,\ldots,r$.
These variables are collected in the column vector $\b Z^{(e)}=(Z_1^{(e)},\ldots,Z_n^{(e)})'$, a generic configuration of which 
is denoted by $\b z=(z_1,\ldots,z_n)'$.
Moreover, let $Y_{ij}^{(e)}$ be a binary variable equal to 1 if units $i$ and $j$ are involved in event $e$, and to 0 otherwise.
Note that
\begin{equation}
Y_{ij}^{(e)}=Z_i^{(e)}Z_j^{(e)},\label{eq:z2y}
\end{equation}
so that the set of variables $Y_{ij}^{(e)}$ is function of the set of variables $Z_i^{(e)}$; the vice-versa does not hold, confirming that the direct analysis of the $Y_{ij}^{(e)}$ leads, in general, to an information loss.
About this point see also the discussion in \cite{aitkin2014statistical}.
\subsection{Marginal effects}
The main issue is how to parametrize the distribution of the random vectors $\b Z^{(e)}$.
We adopt a marginal parametrization  \citep{bergsma2002marginal,bergsma2009marginal} based on hierarchical effects up to a certain order.
This parametrization is less common than the log-linear parametrization, adopted even in ERGMs \citep{frank1986markov,wasserman1996logit}, in which
\[
\log \frac{p(\b Z^{(e)}=\b z)}{p(\b Z^{(e)}=\b 0)}=\b g(\b z)\tr\b\ga,
\]
for all configurations $\b z$ different from the null configuration $\b 0$, where $\b g(\b z)$ is a vector-valued function depending on $\b z$.

Indeed, a marginal parametrization may be expressed on the basis of a sequence of log-linear parametrizations referred to the marginal distribution of selected subset of variables, that is, 
\begin{equation}
\log \frac{p(\b Z_{\cgl M}^{(e)}=\b z_{\cgl M})}{p(\b Z_{\cgl M}^{(e)}=\b 0_{\cgl M})}=\b g_{\cgl M}(\b z_{\cgl M})\tr\b\ga_{\cgl M},\label{eq:marg_par}
\end{equation}
where $\cg M$ is the set of indices of such variables and $\b Z_{\cgl M}^{(e)}$ is the corresponding subvector of $\b Z^{(e)}$.

In our approach, in particular, we rely on first- and second-order effects, specified, for all $e$, as
\ba
\eta_i^{(e)}:=\log\frac{p(Z_i^{(e)}=1)}{p(Z_i^{(e)}=0)},\quad i=1,\ldots,n, \label{Eq:marginal1}
\ea
which is a particular case of (\ref{eq:marg_par}) with $\cg M=\{i\}$, and
\ba
\eta_{ij}^{(e)}:=\log\frac{p(Z^{(e)}_i=0,Z^{(e)}_j=0)p(Z^{(e)}_i=1,Z^{(e)}_j=1)}
{p(Z^{(e)}_i=0,Z^{(e)}_j=1)p(Z^{(e)}_i=1,Z^{(e)}_j=0)},\quad i,j=1,\ldots,n,\: 
j\neq i,
\label{Eq:bivariate1}
\ea
which is obtained from (\ref{eq:marg_par}) with $\cg M=\{i,j\}$.
In terms of interpretation \cite[see also][]{bartolucci2007extended}, we can easily realize that the marginal logit $\eta_i^{(e)}$ is a {\em measure of tendency of unit $i$ to be involved in the $e$-th relational event}.
On the other hand, the log-odds ratio $\eta_{ij}^{(e)}$ is a {\em measure of the tendency of units $i$ and $j$ to cooperate with reference to the same $e$-th relational event}. 

To better interpret the $\eta_{ij}^{(e)}$ effects, it is worth recalling that the log-odds ratio is a well-known measure of association between binary variables \citep[][Ch. 2]{agresti:2013}, being 0 in the case of independence.
In fact, an alternative expression for this effect is
\begin{eqnarray}
\eta_{ij}^{(e)}&=&\log\frac{p(Z^{(e)}_i=1|Z^{(e)}_j=1)}{p(Z^{(e)}_i=0|Z^{(e)}_j=1)}-
\log\frac{p(Z^{(e)}_i=1|Z^{(e)}_j=0)}{p(Z^{(e)}_i=0|Z^{(e)}_j=0)}\nonumber\\
&=&\log\frac{p(Z^{(e)}_j=1|Z^{(e)}_i=1)}{p(Z^{(e)}_j=0|Z^{(e)}_i=1)}-
\log\frac{p(Z^{(e)}_j=1|Z^{(e)}_i=0)}{p(Z^{(e)}_j=0|Z^{(e)}_i=0)},\label{eq:interpret_log-odds}
\end{eqnarray}
corresponding to the increase in the logit of the probability that unit $i$ (or $j$) is involved in the 
$e$-th event, given that unit $j$ (or $i$) is present in the same event, with respect to the case the latter is not present.
More details in this regard are provided in the following section.

Before illustrating how we parametrize in a parsimonious way the marginal effects defined above, it is worth recalling that, apart from the trivial case of $n=2$ actors, the knowledge of these effects is not sufficient to obtain univocally the corresponding distribution of the vectors $\b Z^{(e)}$.
To formulate this argument more formally, let $\b p^{(e)}$ denote the vector containing the $2^n$ joint probabilities $p(\b Z^{(e)}=\b z)$ for all possible configurations $\b z$ in lexicographical order.
Also let $\b\eta_1^{(e)}$ be the vector containing the first-order effects $\eta_i^{(e)}$ for $i=1,\ldots,n$ and let $\b\eta_2^{(e)}$ denote the corresponding vector of second-order effects $\eta_{ij}^{(e)}$ for $i=1,\ldots,n-1$ and $j=i+1,\ldots,n$.
It is possible to prove that
\begin{equation}
\b\eta^{(e)}=\begin{pmatrix}\b\eta_1^{(e)}\cr \b\eta_2^{(e)}\end{pmatrix}=\b C \log(\b M\b p^{(e)})\label{eq:marg_par2}
\end{equation}
for a suitably defined matrix of contrasts $\b C$ and a marginalization matrix $\b M$ with elements equal to 0 or 1.
However this relation is not one-to-one, in the sense that it is not possible to obtain a unique probability vector $\b p^{(e)}$ starting from $\b\eta^{(e)}$.

In order to have an invertible parametrization, the structure of higher order effects must be specified.
Just to give the idea, a third-order marginal effect between units $i$, $j$, and $k$ may be defined as
\[
\eta_{ijk}^{(e)}=\eta_{ij}^{(e)}(Z_k^{(e)}=1)-\eta_{ij}^{(e)}(Z_k^{(e)}=0),
\]
where
\[
\eta_{ij}^{(e)}(Z_k^{(e)}=z)=\log\frac{p(Z^{(e)}_i=0,Z^{(e)}_j=0|Z_k^{(e)}=z)p(Z^{(e)}_i=1,Z^{(e)}_j=1|Z_k^{(e)}=z)}
{p(Z^{(e)}_i=0,Z^{(e)}_j=1|Z_k^{(e)}=z)p(Z^{(e)}_i=1,Z^{(e)}_j=0|Z_k^{(e)}=z)},\quad z=0,1.
\]
This is the difference between the conditional log-odds ratio for units $i$ and $j$ given that unit $k$ is present with respect to the case it is not present.
This directly compares to the {\em triangularization effect} in an ERGM, as it measures how much the presence of unit $k$ affects the chance that units $i$ and $j$ collaborate.
In a similar way we may recursively define effects of order higher than three until order $n$ \citep{bartolucci2007extended}, so that including the specification of these effects, the parametrization in (\ref{eq:marg_par2}) becomes invertible.

In the present approach, however, we prefer to focus only on first- and second-order effects as formulated in (\ref{Eq:marginal1}) and (\ref{Eq:bivariate1}), leaving the structure of higher-order interactions unspecified.
In fact, as we show in detail in Section \ref{sec:inference}, to make inference on these effects it is not necessary to specify the structure of the higher-order effects as we base inference on a pairwise likelihood function.
This is an advantage of the marginal parametrization with respect to the log-linear parametrization; the latter does not allow us to directly express the marginal distribution of a subset of variables without specifying the full set of interactions.
The way of obtaining each bivariate probability vector
\begin{equation}
\b p_{ij}^{(e)}=\begin{pmatrix}
p(Z_i^{(e)}=0,Z_j^{(e)}=0)\cr
p(Z_i^{(e)}=0,Z_j^{(e)}=1)\cr
p(Z_i^{(e)}=1,Z_j^{(e)}=0)\cr
p(Z_i^{(e)}=1,Z_j^{(e)}=1)
\end{pmatrix}
\label{eq:biv_vector}
\end{equation}
on the basis of the parameters $\eta_{i}^{(e)}$, $\eta_j^{(e)}$, and $\eta_{ij}^{(e)}$, which are collected in the vector $\b\eta_{ij}^{(e)}=(\eta_{i}^{(e)},\eta_j^{(e)},\eta_{ij}^{(e)})\tr$, is clarified in the Appendix.
\subsection{Interpretation of the log-odds ratios}\label{sec:log-odds}
To clarify the interpretation of the log-odds ratio $\eta_{ij}^{(e)}$, it is useful to consider that it directly compares with the logit of the probability that there is a connection between units $i$ and $j$, in the sense that both units are involved in the same event.
In fact, from (\ref{eq:z2y}) we have
\[
\tilde{\eta}_{ij}^{(e)}:=\log \frac{p(Y_{ij}^{(e)}=1)}{p(Y_{ij}^{(e)}=0)}=\log \frac{p(Z_i^{(e)}=1,Z_j^{(e)}=1)}{1-p(Z_i^{(e)}=1,Z_j^{(e)}=1)},
\]
that is a commonly used effect in typical social network models; see, for instance, \cite{hoff2002latent}.
There is an important difference between $\tilde{\eta}_{ij}^{(e)}$ and $\eta_{ij}^{(e)}$: the former corresponds to the tendency of units $i$ and $j$ to be involved in the same event, but it does not disentangle this joint tendency from the marginal tendency of each single unit to be involved in the same event.
For instance, $\tilde{\eta}_{ij}^{(e)}$ could attain a large value only because both units have, separately, a high tendency to be involved in the event (both authors are very active in their publication strategy) even if there is not a particular ``attraction'' between them, namely with high values of $p(Z_i^{(e)}=1,Z_j^{(e)}=0)$ or $p(Z_i^{(e)}=0,Z_j^{(e)}=1)$. 
On the other hand, $\eta_{ij}^{(e)}$ is a proper {\em measure of attraction} because, as clearly shown by (\ref{eq:interpret_log-odds}), it corresponds to the increase in the chance that one unit is present 
in the event given that also the other unit is present in the same event.
This difference between parameters $\eta_{ij}^{(e)}$ and $\tilde{\eta}_{ij}^{(e)}$ is key in the proposed approach and is of particular relevance in the application of our interest, where each event may involve a variable number of actors and, possibly, also only one actor.
Indeed, in our approach the general tendency of unit $i$ to be involved in an event is meaningfully measured by effects $\eta_i^{(e)}$ defined in (\ref{Eq:marginal1}).
Note that effects of this type cannot be directly included in an ERGM.

\begin{figure}[ht]\centering
\includegraphics[width=15cm]{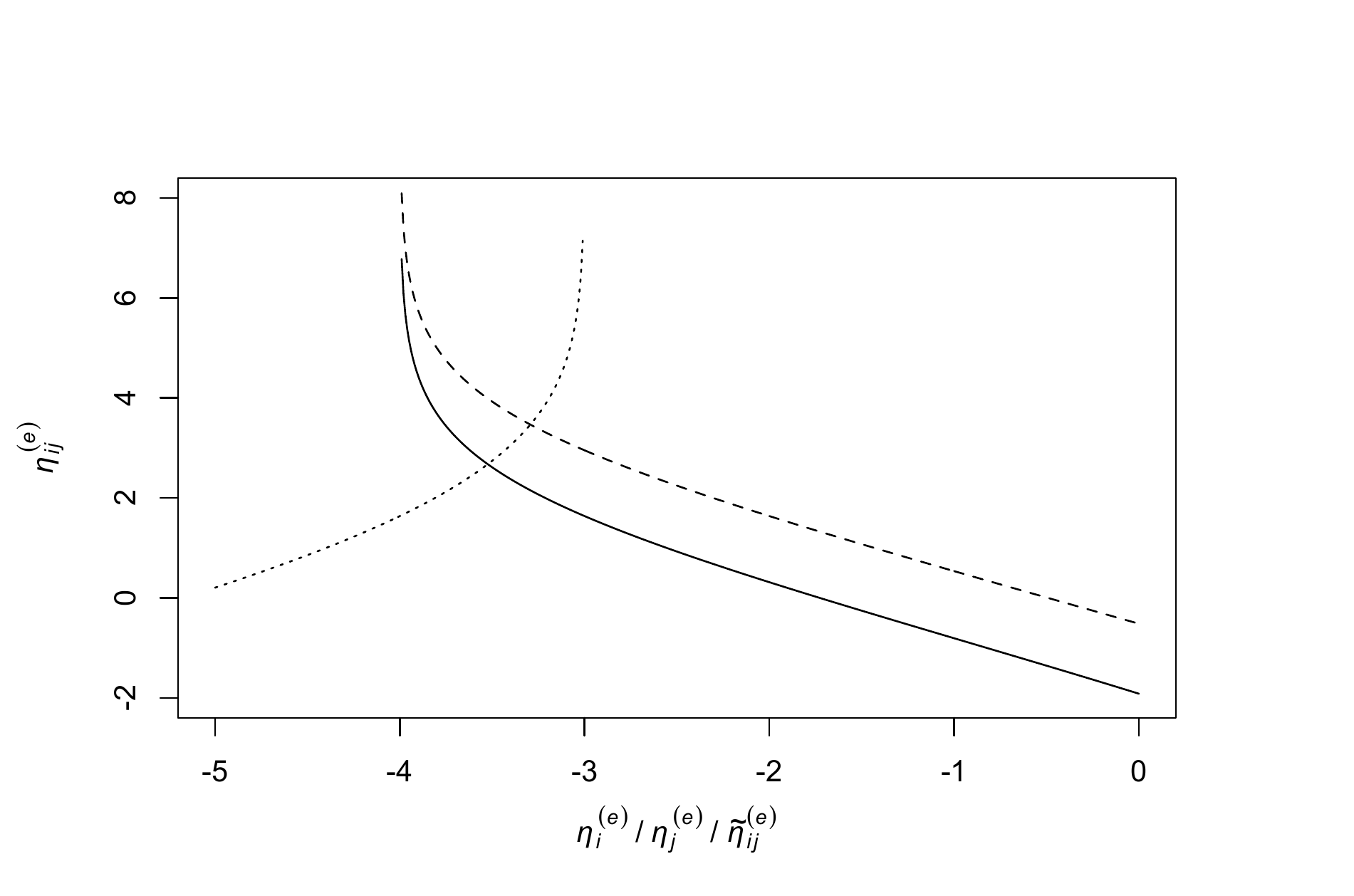}
\caption{\em Plot of the log-odds ratio $\eta_{ij}^{(e)}$ with respect to $\eta_i^{(e)}$ (solid line), $\eta_i^{(e)}$ (dashed line), and $\tilde{\eta}_{ij}^{(e)}$ (dotted line)}\label{fig:log_odds}
\end{figure}

The above arguments may be further clarified considering that, for given marginal distributions $p(Z_i^{(e)})$ and $p(Z_j^{(e)})$, or, in other terms, for fixed $\eta_i^{(e)}$ and $\eta_j^{(e)}$, the log-odds ratio $\eta_{ij}^{(e)}$ is an increasing function of $p(Z_i^{(e)}=1,Z_j^{(e)}=1)=p(Y_{ij}^{(e)}=1)$ and thus of $\tilde{\eta}_{ij}^{(e)}$.
Moreover, for a given value of this joint probability, $\eta_{ij}^{(e)}$ is a decreasing function of $\eta_i^{(e)}$ and $\eta_j^{(e)}$.
In particular, considering that
\begin{eqnarray*}
p(Z_i^{(e)}=0,Z_j^{(e)}=0)&=&1-p(Z_i^{(e)}=1)-p(Z_j^{(e)}=1)+p(Z_i^{(e)}=1,Z_j^{(e)}=1),\\
p(Z_i^{(e)}=0,Z_j^{(e)}=1)&=&p(Z_j^{(e)}=1)-p(Z_i^{(e)}=1,Z_j^{(e)}=1),\\
p(Z_i^{(e)}=1,Z_j^{(e)}=0)&=&p(Z_i^{(e)}=1)-p(Z_i^{(e)}=1,Z_j^{(e)}=1),
\end{eqnarray*}
we can easily realize that
\[
\frac{\pa\eta_{ij}^{(e)}}{\pa\tilde{\eta}_{ij}^{(e)}}=p(Z_i^{(e)}=1,Z_i^{(e)}=1)[1-p(Z_i^{(e)}=1,Z_i^{(e)}=1)]\sum_{z_1=0}^1\sum_{z_2=0}^1\frac{1}{p(Z_i^{(e)}=z_1,Z_i^{(e)}=z_2)}>0.
\]
Similarly, we have
\[
\frac{\pa\eta_{ij}^{(e)}}{\pa\eta_i^{(e)}}=-p(Z_i^{(e)}=1)[1-p(Z_i^{(e)}=1)]\sum_{z_1=0}^1\sum_{z_2=0}^1\frac{1}{p(Z_i^{(e)}=z_1,Z_i^{(e)}=z_2)}<0,
\]
with a corresponding expression for $\pa\eta_{ij}^{(e)}/\pa\eta_j^{(e)}$.
An illustration of this behavior is provided in Figure \ref{fig:log_odds}, where for a pair of individuals we represent the value of $\eta_{ij}^{(e)}$ with respect to $\eta_i^{(e)}$ (with $\eta_j^{(e)}=-2$ and $\tilde\eta_{ij}^{(e)}=-4$), to $\eta_j^{(e)}$ (with $\eta_i^{(e)}=-3$ and $\tilde\eta_{ij}^{(e)}=-4$), and to $\tilde\eta_{ij}^{(e)}$ (with $\eta_i^{(e)}=-3$ and $\eta_j^{(e)}=-2$).

Another advantage of $\eta_{ij}^{(e)}$ with respect to $\tilde{\eta}_{ij}^{(e)}$ is that the former induces a variational independent parametrization \citep{bergsma2002variation}.
This means that the joint distribution of $(Z_i^{(e)},Z_j^{(e)})'$ exists for any value in $\mathbb{R}$ of the first-order effects $\eta_i^{(e)}$ and $\eta_j^{(e)}$ and of the second-order effect $\eta_{ij}^{(e)}$.
More formally, the function relating $\b\eta_{ij}^{(e)} = (\eta_{i}^{(e)},\eta_{j}^{(e)},\eta_{ij}^{(e)})'$ with $\b p_{ij}^{(e)}$ defined in (\ref{eq:biv_vector}) is one-to-one for $\b\eta_{ij}^{(e)}$ in $\mathbb{R}^3$ and $\b p_{ij}^{(e)}$ in the four-dimensional simplex.
This has advantages in terms of model interpretation and estimation.
On the other hand, effect $\tilde{\eta}_{ij}^{(e)}$ has a limited range of possible values with bounds depending on $\eta_i^{(e)}$ and $\eta_j^{(e)}$, making the joint estimation and interpretation of $\tilde{\eta}_{ij}^{(e)}, \eta_i^{(e)}$ and $\eta_j^{(e)}$, more problematic.
\subsection{Parametrization of marginal effects}
Formulating a model for relational events requires to parametrize, in a parsimonious way, the effects $\eta_i^{(e)}$ and $\eta_{ij}^{(e)}$ of main interest.
In absence of individual covariates, we propose the following parametrization of the first-order effects:
\begin{equation}
\eta_i^{(e)}=\b f_1(t_e)\tr\b\al_i,\label{eq:par1}
\end{equation}
where $\b f_1(t_e)$ is a vector-valued function specific of time $t_e$ of each event $e$.
For instance, this function may contain the terms of a polynomial of suitable order of the day or year of event $e$ starting from the beginning of the study.
This parametrization is similar to that of a latent trajectory model \citep{dwye:1983,crow:hand:1996,mena:2002}, with the difference that, as we clarify in the sequel, each $\b\al_i$ is considered as a vector of fixed individual parameters.  
In any case, it is possible to represent individual trajectories regarding the tendency over time to be present in an event.

Regarding the second-order effects, a natural extension of (\ref{eq:par1}) would lead to a vector of specific parameters for each pair of units.
However, to obtain a parsimonious model, we prefer to rely on an additive parametrization of type
\begin{equation}
\eta_{ij}^{(e)}=\b f_2(t_e)\tr(\b\be_i+\b\be_j),\label{eq:par2}
\end{equation}
where $\b f_2(t_e)$ is defined as $\b f_1(t_e)$ and, again, vectors $\b\be_i$ represent the evolution of the tendency, of unit $i$, to collaborate across time.
We use two different functions, $\b f_1(t_e)$ and $\b f_2(t_e)$, to allow for a different order of the involved polynomials of time.
Note, however, that the additive structure in (\ref{eq:par2}) implies that $\b f_2(t_e)\tr\b\be_i$ may be interpreted as the ``general'' tendency of individual $i$ to collaborate with other individuals in an event at time $t_e$.

Overall, for each bivariate probability vector $\b p_{ij}^{(e)}$, the parametrization based on (\ref{eq:par1}) and (\ref{eq:par2}) is linear in the parameters.
In particular, if we let $\b\de_i=(\b\al_i\tr,\b\be_i\tr)\tr$, we have that
\begin{equation}
\b\eta_{ij}^{(e)}=\begin{pmatrix}
\b D_{ij1} & \b D_{ij2}
\end{pmatrix}
\begin{pmatrix}
\b\de_i\cr \b\de_j
\end{pmatrix},\label{eq:delta}
\end{equation}
where $\b D_{ij1}$ and $\b D_{ij2}$ are suitable design matrices.

To clarify the proposed parametrization, consider a sample of $n=9$ individuals for a single event, for different values of the intercepts $\al_i$ (from -3 to -1) and of $\be_i$ (from -1 to 1).
These values are reported in Table \ref{tab:al_be} together with certain average probabilities that help to understand the meaning of these parameters.
The single $2\times 2$ tables for each pair of actors are reported in Table \ref{tab:2x2}.

\begin{table}[ht]
\centering
\begin{tabular}{rrrrr}
  \hline\hline
 & $\al_i$ & $\be_i$ & mean($p(Z_i^{(e)})$) &  mean($p(Z_i^{(e)},Z_j^{(e)}=1)$) \\ 
  \hline
1 & -3.00 & -1.00 & 0.05 & 0.00 \\ 
  2 & -3.00 & 0.00 & 0.05 & 0.01 \\ 
  3 & -3.00 & 1.00 & 0.05 & 0.01 \\ 
  4 & -2.00 & -1.00 & 0.12 & 0.01 \\ 
  5 & -2.00 & 0.00 & 0.12 & 0.02 \\ 
  6 & -2.00 & 1.00 & 0.12 & 0.03 \\ 
  7 & -1.00 & -1.00 & 0.27 & 0.02 \\ 
  8 & -1.00 & 0.00 & 0.27 & 0.04 \\ 
  9 & -1.00 & 1.00 & 0.27 & 0.05 \\ 
   \hline
\end{tabular}
\caption{\em Parameters $\al_i$ and $\be_i$ together with mean marginal probabilities and probabilities of two actors being involved in the same event}\label{tab:al_be}
\end{table}

\begin{sidewaystable}[ht]
\centering
{\footnotesize
\begin{tabular}{|rr|rr|rr|rr|rr|rr|rr|rr|rr|rr|}
  \hline
& & \multicolumn2{c|}{$j=1$}  & \multicolumn2{c|}{$j=2$}
 & \multicolumn2{c|}{$j=3$}  & \multicolumn2{c|}{$j=4$}
 & \multicolumn2{c|}{$j=5$}  & \multicolumn2{c|}{$j=6$}
 & \multicolumn2{c|}{$j=7$}  & \multicolumn2{c|}{$j=8$}
 & \multicolumn2{c|}{$j=9$} \\
& & \multicolumn1c{0} & \multicolumn1{c|}{1}
 & \multicolumn1c{0} & \multicolumn1{c|}{1}
 & \multicolumn1c{0} & \multicolumn1{c|}{1}
 & \multicolumn1c{0} & \multicolumn1{c|}{1}
 & \multicolumn1c{0} & \multicolumn1{c|}{1}
 & \multicolumn1c{0} & \multicolumn1{c|}{1}
 & \multicolumn1c{0} & \multicolumn1{c|}{1}
 & \multicolumn1c{0} & \multicolumn1{c|}{1}
 & \multicolumn1c{0} & \multicolumn1{c|}{1}  \\ 
  \hline
\multirow{2}{*}{$i=1$} &   
0 & - & - & 0.906 & 0.047 & 0.907 & 0.045 & 0.834 & 0.118 & 0.836 & 0.117 & 0.839 & 0.114 & 0.686 & 0.267 & 0.689 & 0.263 & 0.696 & 0.256 \\ 
& 1 & - & - & 0.047 & 0.001 & 0.045 & 0.002 & 0.047 & 0.001 & 0.045 & 0.002 & 0.042 & 0.006 & 0.045 & 0.002 & 0.042 & 0.006 & 0.035 & 0.013 \\\hline 
\multirow{2}{*}{$i=2$} & 0 & 0.906 & 0.047 & - & - & 0.910 & 0.042 & 0.836 & 0.117 & 0.839 & 0.114 & 0.846 & 0.107 & 0.689 & 0.263 & 0.696 & 0.256 & 0.707 & 0.246 \\ 
& 1 & 0.047 & 0.001 & - & - & 0.042 & 0.005 & 0.045 & 0.002 & 0.042 & 0.006 & 0.035 & 0.012 & 0.042 & 0.006 & 0.035 & 0.013 & 0.024 & 0.023 \\\hline
\multirow{2}{*}{$i=3$} & 0 & 0.907 & 0.045 & 0.910 & 0.042 & - & - & 0.839 & 0.114 & 0.846 & 0.107 & 0.855 & 0.098 & 0.696 & 0.256 & 0.707 & 0.246 & 0.717 & 0.235 \\ 
& 1 & 0.045 & 0.002 & 0.042 & 0.005 & - & - & 0.042 & 0.006 & 0.035 & 0.012 & 0.026 & 0.022 & 0.035 & 0.013 & 0.024 & 0.023 & 0.014 & 0.034 \\\hline 
\multirow{2}{*}{$i=4$} & 0 & 0.834 & 0.047 & 0.836 & 0.045 & 0.839 & 0.042 & - & - & 0.768 & 0.113 & 0.776 & 0.105 & 0.618 & 0.262 & 0.627 & 0.254 & 0.644 & 0.237 \\ 
& 1 & 0.118 & 0.001 & 0.117 & 0.002 & 0.114 & 0.006 & - & - & 0.113 & 0.006 & 0.105 & 0.014 & 0.113 & 0.006 & 0.104 & 0.015 & 0.087 & 0.032 \\\hline 
\multirow{2}{*}{$i=5$} & 0 & 0.836 & 0.045 & 0.839 & 0.042 & 0.846 & 0.035 & 0.768 & 0.113 & - & - & 0.790 & 0.091 & 0.627 & 0.254 & 0.644 & 0.237 & 0.667 & 0.213 \\ 
& 1 & 0.117 & 0.002 & 0.114 & 0.006 & 0.107 & 0.012 & 0.113 & 0.006 & - & - & 0.091 & 0.028 & 0.104 & 0.015 & 0.087 & 0.032 & 0.064 & 0.055 \\\hline 
\multirow{2}{*}{$i=6$} & 0 & 0.839 & 0.042 & 0.846 & 0.035 & 0.855 & 0.026 & 0.776 & 0.105 & 0.790 & 0.091 & - & - & 0.644 & 0.237 & 0.667 & 0.213 & 0.692 & 0.189 \\ 
& 1 & 0.114 & 0.006 & 0.107 & 0.012 & 0.098 & 0.022 & 0.105 & 0.014 & 0.091 & 0.028 & - & - & 0.087 & 0.032 & 0.064 & 0.055 & 0.039 & 0.080 \\\hline 
\multirow{2}{*}{$i=7$} & 0 & 0.686 & 0.045 & 0.689 & 0.042 & 0.696 & 0.035 & 0.618 & 0.113 & 0.627 & 0.104 & 0.644 & 0.087 & - & - & 0.501 & 0.230 & 0.534 & 0.197 \\ 
& 1 & 0.267 & 0.002 & 0.263 & 0.006 & 0.256 & 0.013 & 0.262 & 0.006 & 0.254 & 0.015 & 0.237 & 0.032 & - & - & 0.230 & 0.039 & 0.197 & 0.072 \\\hline 
\multirow{2}{*}{$i=8$} & 0 & 0.689 & 0.042 & 0.696 & 0.035 & 0.707 & 0.024 & 0.627 & 0.104 & 0.644 & 0.087 & 0.667 & 0.064 & 0.501 & 0.230 & - & - & 0.576 & 0.155 \\  & 1 & 0.263 & 0.006 & 0.256 & 0.013 & 0.246 & 0.023 & 0.254 & 0.015 & 0.237 & 0.032 & 0.213 & 0.055 & 0.230 & 0.039 & - & - & 0.155 & 0.114 \\\hline 
\multirow{2}{*}{$i=9$} & 0 & 0.696 & 0.035 & 0.707 & 0.024 & 0.717 & 0.014 & 0.644 & 0.087 & 0.667 & 0.064 & 0.692 & 0.039 & 0.534 & 0.197 & 0.576 & 0.155 & - & - \\ 
& 1 & 0.256 & 0.013 & 0.246 & 0.023 & 0.235 & 0.034 & 0.237 & 0.032 & 0.213 & 0.055 & 0.189 & 0.080 & 0.197 & 0.072 & 0.155 & 0.114 & - & - \\\hline 
\end{tabular}}
\caption{\em Single $2\times 2$ tables for each pair of actors considered in Table \ref{tab:al_be}}\label{tab:2x2}
\end{sidewaystable}
\section{Pairwise likelihood inference}\label{sec:inference}
Before introducing the proposed methods of inference for the model described above, we clarify the data structure used in applications.
We start from the data
\begin{equation}
\b w_{ij}^{(e)}=\begin{pmatrix}
I(z_i^{(e)}=0,z_j^{(e)}=0)\cr
I(z_i^{(e)}=0,z_j^{(e)}=1)\cr
I(z_i^{(e)}=1,z_j^{(e)}=0)\cr
I(z_i^{(e)}=1,z_j^{(e)}=1)\cr
\end{pmatrix},\label{eq:freq_vector}
\end{equation}
where $I(\cdot)$ is the indicator function, for $i=1,\ldots,n-1$, $j=i+1,\ldots,n$, and $e=1,\ldots,r$.
Moreover, in the applications of interest, it is possible to group events that, by assumption, have the same distribution.
For instance, in the application based on the academic articles published by statisticians, it is plausible to assume that for all articles published in the same year the distribution of the binary vector is the same, even because it is not possible to have the precise dates of the publication and 
thus their precise time order.
In other words, it is sensible to group events in homogenous periods $t=1,\ldots,t(r)$, where $t(e)$ denotes the time of event $e$.
Then, the relevant information is that contained in the frequency vectors
\[
\tilde{\b w}_{ij}^{(t)}=\sum_{e:t(e)=t}\b w_{ij}^{(e)},
\]
and consequently we denote by $\tilde{\b p}_{ij}^{(t)}$ the corresponding probability vector having the same structure as in (\ref{eq:biv_vector}).
\subsection{Fixed-effects estimation}\label{sec:fe_estimation}
It is possible to estimate the parameters of interest by maximizing the pairwise log-likelihood function \citep{lindsay1988composite,varin2011overview}:
\[
p\ell(\b\th)=\sum_{i=1}^{n-1}\sum_{j=i+1}^n\sum_{e=1}^r[\b w_{ij}^{(e)}]\tr\log\b p_{ij}^{(e)},
\]
where $\b\th$ is the vector of all such parameters, that is the collection of individual parameter vectors $\b\de_i$, $i=1,\ldots,n$, used in (\ref{eq:delta}).
An alternative expression is
\begin{eqnarray}
p\ell(\b\th)&=&
\sum_{i=1}^{n-1}\sum_{j=i+1}^n\ell_{ij}(\b\de_i,\b\de_j),\label{Eq:loglik}\\
\ell_{ij}(\b\de_i,\b\de_j)&=&\sum_{t=1}^{t(r)}[\tilde{\b w}_{ij}^{(t)}]\tr\log\tilde{\b p}_{ij}^{(t)},\nonumber 
\end{eqnarray}
which is faster to compute as it relies on the frequency vectors defined in (\ref{eq:freq_vector}).

In order to maximize $p\ell(\b\th)$, it is important to obtain the score vector of each components $\ell_{ij}(\b\de_i,\b\de_j)$.
To this aim, it is convenient to introduce the log-linear effects $\tilde{\b p}_{ij}^{(t)}$ which are collected in the vector $\tilde{\b\la}_{ij}^{(t)} = (\la_i^{(e)},\la_j^{(e)},\la_{ij}^{(e)})\tr$, where 
\begin{eqnarray*}
\la_i^{(e)}&=&\log\frac{p(Z_i^{(e)}=1|Z_j^{(e)}=0)}{p(Z_i^{(e)}=0|Z_j^{(e)}=0)},\\
\la_j^{(e)}&=&\log\frac{p(Z_j^{(e)}=1|Z_i^{(e)}=0)}{p(Z_j^{(e)}=0|Z_i^{(e)}=0)},\\
\la_{ij}^{(e)}&=&\eta_{ij}^{(e)},
\end{eqnarray*}
and $e$ is any of the events at time occasion $t$.
Also let $\tilde{\b\eta}_{ij}^{(t)}$ denote the corresponding vector of marginal parameters.
We have that
\[
\b s_{ij}(\b\de_i):=\frac{\pa\ell_{ij}(\b\de_i,\b\de_j)}{\pa\b\de_i}=\sum_{t=1}^{t(r)}\b D_{ij1}\tr\frac{\b\pa\b\eta_{ij}^{(t)}}{\pa[\b\la_{ij}^{(t)}]\tr}\b G[\tilde{\b w}_{ij}^{(t)}-m_{ij}^{(t)}\tilde{\b p}_{ij}^{(t)}],
\]
where $m_{ij}^{(t)}$ is the sum of the elements of $\tilde{\b w}_{ij}^{(t)}$, that is, the number of events in time period $t$, whereas $\b G$ and the derivative of $\tilde{\b\eta}_{ij}^{(t)}$ with respect to $\tilde{\b\la}_{ij}^{(t)}$ are defined in Appendix.

The estimation algorithm is based on the following steps.
First of all define an initial guess for the parameters $\b\de_i$, denoted by $\b\de_i^{(0)}$, $i=1,\ldots,n$.
Then, for every unit $i$, find the values of $\b\de_i$ such that
\[
\sum_{j=1,\:j\neq i}^n\b s_{ij}(\b\de_i) = \b 0,
\]
so as to maximize
\begin{equation}
p\ell_i(\b\th)=\sum_{j=1,\:j\neq i}^n\ell_{ij}(\b\de_i,\b\de_j),\label{eq:individual_pl}
\end{equation}
with respect to $\b\de_i$, with all other parameters kept fixed. Iterate this process until convergence in $p\ell(\b\th)$, and denote the final parameter estimates by $\hat{\b\de}_i=(\hat{\b\al}_i',\hat{\b\be}_i')'$, $i=1,\ldots,n$, which are collected in the vector $\hat{\b\th}$. 
In practice, the algorithm steps may be implemented by using a readily available numerical solver.
\subsection{Clustering}\label{sec:clustering}
With large samples, it is typically of interest to find clusters of units presenting a similar behavior.
In our approach this amounts to assume that there are $h_1$ groups of individuals having a similar behavior in terms of tendency to be involved in an event and $h_2$ groups of individuals having a similar tendency to collaborate in the network.
For each group we have specific parameter vectors denoted by $\b\al_{g_1}^*$ and $\b\be_{g_2}^*$, with $g_1=1,\ldots,h_1$ and $g_2=1,\ldots,h_2$, all collected in the parameter vector $\b\th^*$.

For unit $i$, let $d_{i1}$ denote the cluster to which the unit is assigned with respect to the first type of tendency and $d_{i2}$ the cluster assigned with respect to the second type of tendency.
The corresponding classification pairwise log-likelihood has the same expression as $p\ell(\b\th)$ defined in (\ref{Eq:loglik}), with $\b\al_i=\b\al_{d_{i1}}^*$, $\b\be_i=\b\be_{d_{i2}}^*$, and then $\b\de_i=((\b\al_{d_{i1}}^*)\tr,(\b\be_{d_{i2}}^*)\tr)\tr$.
This function is denoted by $cp\ell(\b\th^*,\b d_1,\b d_2)$, where $\b d_1$ is the vector with elements $d_{i1}$ and $\b d_2$ is that with elements $d_{i2}$, respectively, with $i=1,\ldots,n$.

To cluster units in homogeneous groups, we maximize $cp\ell(\b\th,\b d_1,\b d_2)$ by an iterative algorithm that is initialized from the output of a $k$-means clustering of the individual estimates $\hat{\b\al}_i$ and $\hat{\b\be}_i$. 
Then, it alternates the following three   steps until convergence:
\begin{enumerate}
\item for $i=1,\ldots,n$ try to change the cluster $d_{i1}$ of unit $i$ by finding the cluster that maximizes the individual component of the classification pairwise log-likelihood, which is defined as in (\ref{eq:individual_pl}) accounting for the cluster structure, with all other parameters kept fixed;
\item for $i=1,\ldots,n$ try to change the cluster $d_{i2}$ of unit $i$ by the same procedure as above;
\item update the parameter estimates of $\b\al_{g_1}^*$, $g_1=1,\ldots,h_1$, and $\b\be_{g_2}^*$,  $g_2=1,\ldots,h_2$, by maximizing $cp\ell(\b\th^*,\b d_1,\b d_2)$ with respect to $\b\th^*$ with $\b d_1$ and $\b d_2$ kept fixed.
\end{enumerate} 

We select $h_1$ and $h_2$ as the smallest number of clusters such that the initial clustering of the estimates $\hat{\b\al}_i$ and $\hat{\b\be}_i$, performed by the $k$-means algorithm, leads to a between sum of squares equal to at least 80\% of the total sum of squares.
Then, at convergence of the three steps illustrated above, we check that the number of clusters is adequate, comparing the maximum value of $cp\ell(\b\th^*,\b d_1,\b d_2)$ with that of $p\ell(\b\th)$, as we show in connection with the application in the next section.
%
\section{Application}\label{sec:app}
\begin{figure}[ht]\centering
\begin{tabular}{cc}
\includegraphics[width=8cm]{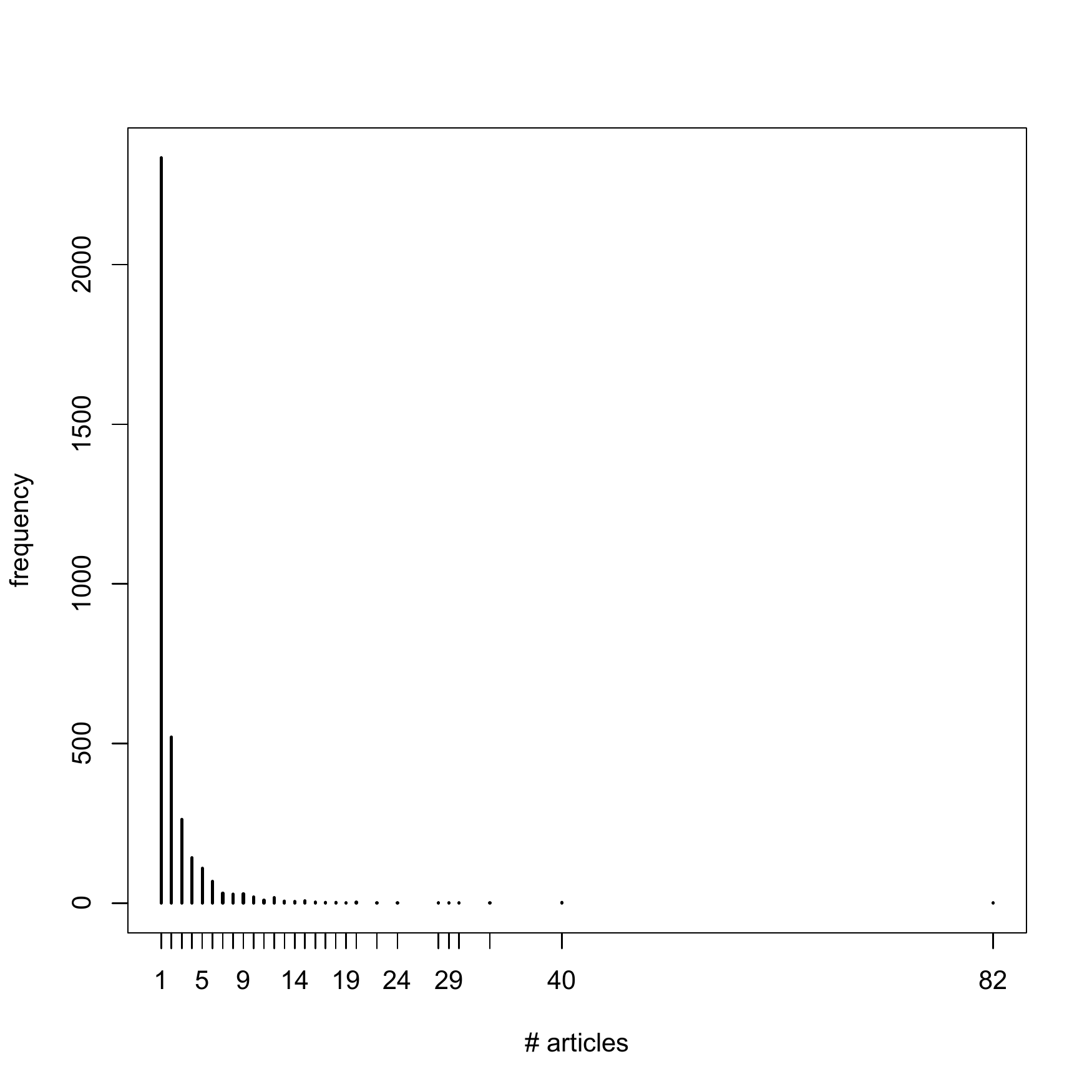} & \includegraphics[width=8cm]{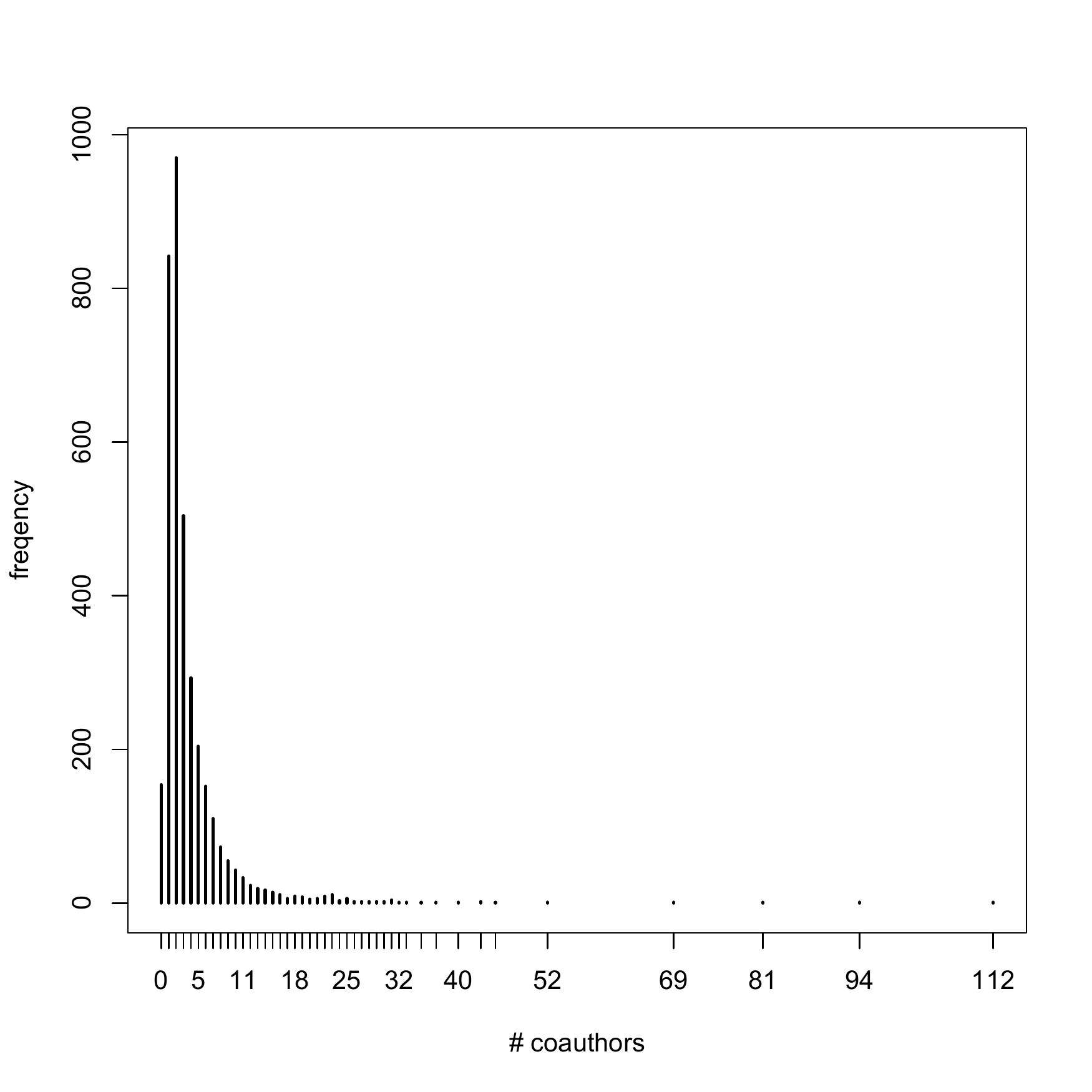}
\end{tabular}
\caption{\em Distribution of the number of articles per author (left panel) and of the number of coauthors (right panel)}\label{fig:descriptive}
\end{figure}

%
In order to illustrate the approach based on individual-specific effects, see assumptions (\ref{eq:par1}) and (\ref{eq:par2}), we propose an application based on the data recently made available by \cite{ji2017coauthorship}.
The data refer to the publication history of all authors with at least one paper published in four top statistical journals ({\em Annals of Statistics}, {\em Biometrika}, {\em Journal of the American Statistical Association}, {\em Journal of the Royal Statistical Society - series B}) between 2003 and the first half of 2012.
Overall, 3,607 authors are involved who coauthored 3,248 articles. 
In Table \ref{tab:descriptive} we report some descriptive statistics, whereas in Figure \ref{fig:descriptive} we represent the distribution of the number of articles and the number of coauthors for each individual in the dataset.

\begin{table}[ht]\centering
\centering
\begin{tabular}{rrrrrrrrrrrrr}
  \hline\hline
  & & \multicolumn{11}c{\# authors}\\\cline{2-13}
year & \# articles & 1 & 2 & 3 & 4 & 5 & 6 & 7 & 8 & 9 & 10 & average \\ 
  \hline
2003 & 296 & 78 & 132 & 63 & 19 & 2 & 2 & 0 & 0 & 0 & 0 & 2.125 \\ 
  2004 & 320 & 70 & 157 & 67 & 17 & 5 & 3 & 1 & 0 & 0 & 0 & 2.197 \\ 
  2005 & 328 & 64 & 166 & 77 & 16 & 4 & 0 & 1 & 0 & 0 & 0 & 2.189 \\ 
  2006 & 354 & 55 & 178 & 96 & 18 & 4 & 2 & 1 & 0 & 0 & 0 & 2.288 \\ 
  2007 & 350 & 56 & 158 & 105 & 24 & 3 & 2 & 1 & 0 & 0 & 1 & 2.363 \\ 
  2008 & 370 & 59 & 151 & 114 & 32 & 9 & 2 & 2 & 1 & 0 & 0 & 2.459 \\ 
  2009 & 409 & 53 & 177 & 128 & 42 & 4 & 2 & 0 & 1 & 1 & 1 & 2.489 \\ 
  2010 & 355 & 53 & 151 & 104 & 36 & 8 & 2 & 1 & 0 & 0 & 0 & 2.451 \\ 
  2011 & 325 & 39 & 135 & 107 & 34 & 5 & 4 & 1 & 0 & 0 & 0 & 2.529 \\ 
  2012 & 141 & 16 & 63 & 49 & 9 & 2 & 1 & 1 & 0 & 0 & 0 & 2.468 \\\hline
& 3248 & 543 & 1468 & 910 & 247 & 46 & 20 & 9 & 2 & 1 & 2 & 2.357 \\ 
   \hline
\end{tabular}
\caption{\em Descriptive statistics on the number of authors per article}\label{tab:descriptive}
\end{table}

From Table \ref{tab:descriptive} we observe that the number of published papers per year does not vary considerably, even if there is an increase from 2003 to 2009 and a decrease after 2009 (year 2012 counts only partially).
Moreover, the average number of authors moderately increases during the time span and it is important to note that the number of single-author articles is relevant in each year.
Indeed, these articles represent the 16.7\% of the total articles considered in the dataset and this justifies the use of the proposed approach for the analysis. Regarding Figure \ref{fig:descriptive}, we note the concentration of the number of articles, with the majority of authors (2,335) who published only one article in one of the four top journals considered in the reference period and 520 who published only two articles.
On the other hand, the three most productive researchers published 33, 40, and 82 articles, with a total of 155 overall (ignoring possible overlapping).
Even the distribution of the number of coauthors shows a very high concentration, although the situation is somehow different as the third modality (i.e., 2 coauthors) has the highest frequency, equal to 970.
The authors with zero and one coauthors are 154 and 842, respectively, whereas the three highest modalities are 81, 94 and 112.

The application is based on two phases, that is, fixed-effects estimation and clustering, which are illustrated in Sections \ref{sec:fe_estimation} and \ref{sec:clustering}, respectively.
Regarding the first phase, with reference to (\ref{eq:par1}) and (\ref{eq:par2}), we assume second order polynomials for the effect of time. 
In this way we estimate 3,607 parameter vectors $\b\al_i$ and $\b\be_i$ of length 3.
On the basis of these estimates it is possible to obtain trajectories both in terms of tendency to publish an article in a certain period and in terms of tendency to collaborate with other authors.
We recall that, for the former, the effect that is represented is the logit defined in (\ref{Eq:marginal1}) and for the latter it is the log-odds ratio (\ref{Eq:bivariate1}).
In order to illustrate these results, we consider the five authors with the largest number of published articles in the period that we identify with the letters from A to E; the patterns of publication of these authors is reported in Table \ref{tab:authors}.

\begin{table}[ht]
\centering
{\small
\begin{tabular}{llrrrrrrrrrrrr}
  \hline\hline
       &       & \multicolumn{10}c{year}\\\cline{3-12}
author &       & 2003 & 2004 & 2005 & 2006 & 2007 & 2008 & 2009 & 2010 & 2011 & 2012 & total \\ 
  \hline
A & \# papers  & 2 & 1 & 6 & 3 & 1 & 4 & 4 & 4 & 3 & 2 & 30 \\ 
  & \# coauth. & 3 & 1 & 8 & 5 & 2 & 5 & 5 & 4 & 7 & 3 & 43 \\ 
  & ratio      & 1.50 & 1.00 & 1.33 & 1.67 & 2.00 & 1.25 & 1.25 & 1.00 & 2.33 & 1.50 & 1.43 \\\hline 
B & \# papers  & 1 & 2 & 3 & 3 & 3 & 3 & 4 & 5 & 9 & 0 & 33 \\ 
  & \# coauth. & 1 & 2 & 3 & 3 & 4 & 6 & 4 & 8 & 12 & 0 & 43 \\ 
  & ratio      & 1.00 & 1.00 & 1.00 & 1.00 & 1.33 & 2.00 & 1.00 & 1.60 & 1.33 & - & 1.30 \\\hline 
C & \# papers  & 2 & 5 & 3 & 1 & 6 & 6 & 6 & 5 & 3 & 3 & 40 \\ 
  & \# coauth. & 3 & 5 & 6 & 2 & 13 & 9 & 10 & 9 & 6 & 6 & 69 \\ 
  & ratio      & 1.50 & 1.00 & 2.00 & 2.00 & 2.17 & 1.50 & 1.67 & 1.80 & 2.00 & 2.00 & 1.73 \\\hline 
D & \# papers  & 2 & 5 & 3 & 3 & 6 & 3 & 8 & 4 & 4 & 2 & 40 \\ 
  & \# coauth. & 6 & 13 & 7 & 3 & 14 & 5 & 17 & 15 & 10 & 4 & 94 \\ 
  & ratio      & 3.00 & 2.60 & 2.33 & 1.00 & 2.33 & 1.67 & 2.12 & 3.75 & 2.50 & 2.00 & 2.35 \\\hline 
E & \# papers  & 7 & 7 & 10 & 12 & 10 & 8 & 11 & 9 & 6 & 2 & 82 \\ 
  & \# coauth. & 7 & 10 & 13 & 16 & 13 & 12 & 16 & 10 & 12 & 3 & 112 \\ 
  & ratio      & 1.00 & 1.43 & 1.30 & 1.33 & 1.30 & 1.50 & 1.45 & 1.11 & 2.00 & 1.50 & 1.37 \\ 
   \hline
\end{tabular}}
\caption{\em Publication profiles of the five authors with the largest number of published papers in the period considered}\label{tab:authors}
\end{table}

\begin{figure}[ht]\centering
\begin{tabular}{cc}
\includegraphics[width=8cm]{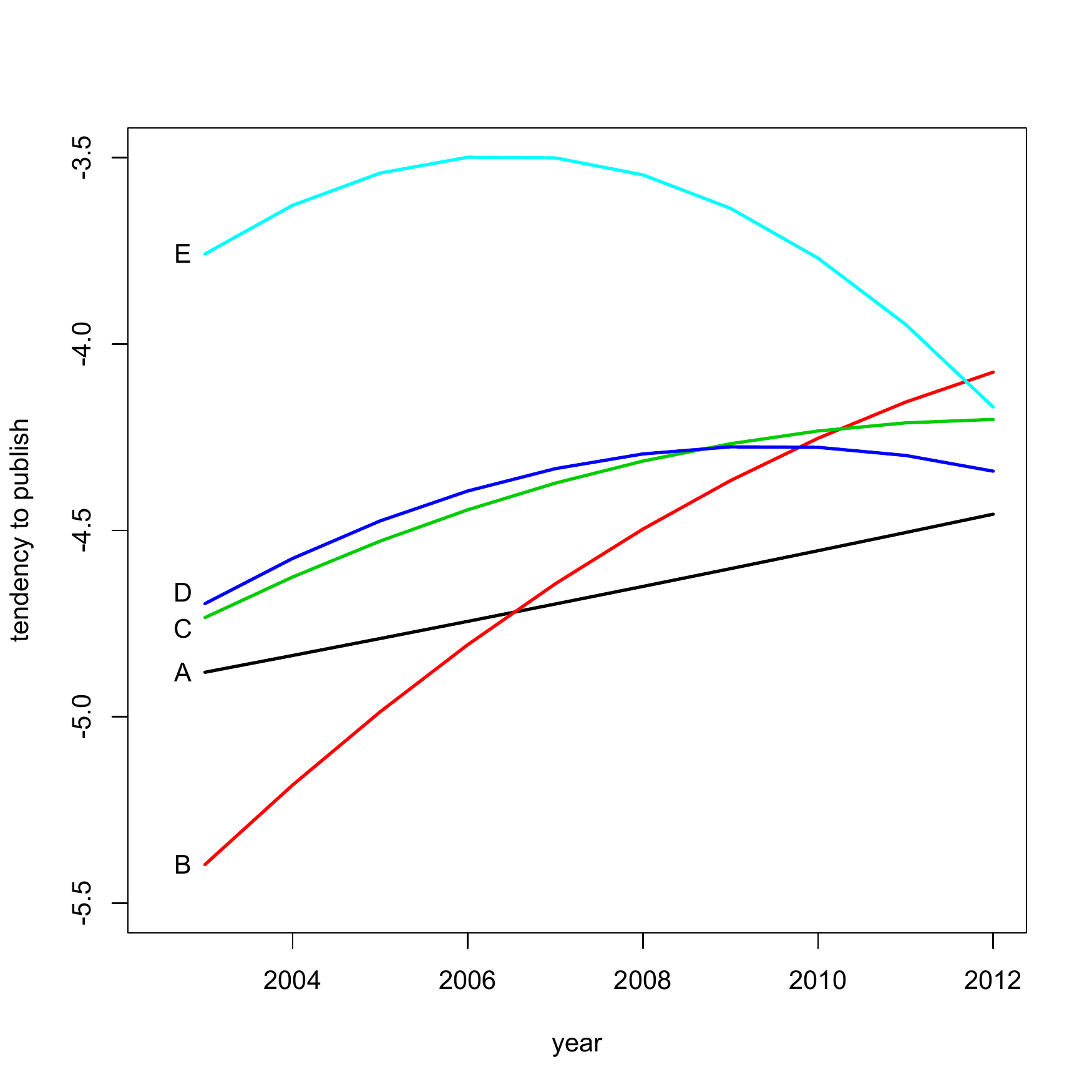} & \includegraphics[width=8cm]{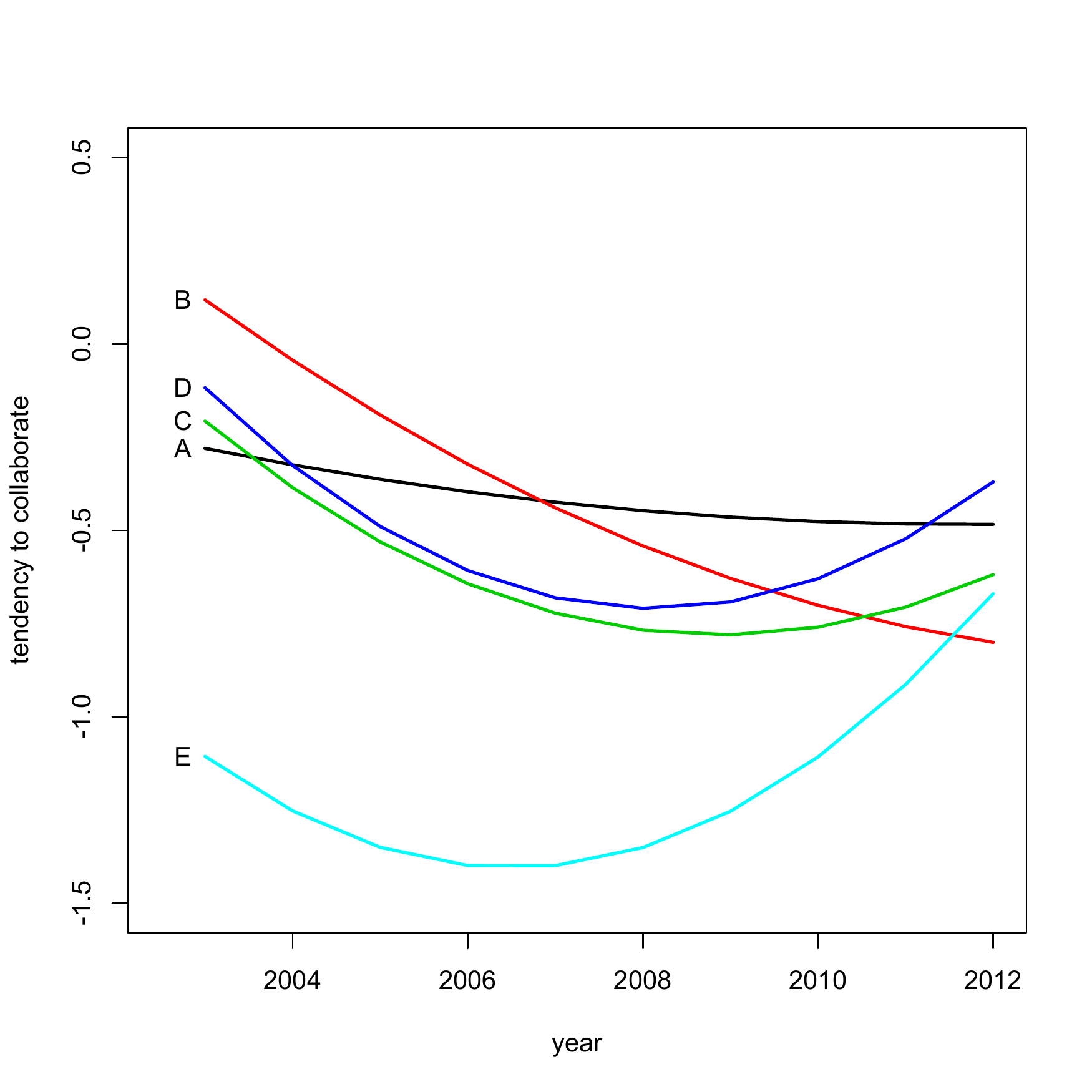}
\end{tabular}
\caption{\em Profiles in terms of tendency of authors to publish papers (left panel) and of collaborate (right panel)}\label{fig:prof_ind}
\end{figure}

For these top five authors, we represent the estimated profiles in Figure \ref{fig:prof_ind} where we can clearly identify author E as the most productive one with a profile that follows a reverse U-shape, having its pick around years 2006 and 2007, coherently with the data in Table \ref{tab:authors}.
On the other hand, it may seem surprising that this author shows the lowest profile in terms of tendency to collaborate with other authors.
However, coherently the data in the table, in terms of ratio between number of coauthors and number of published papers, author E tends to be lower than everybody else.
The conclusion is that the large number of coauthors of author E can be mostly ascribed to his tendency to publish; see Section \ref{sec:log-odds} for general comments on the interpretation of these results.
In a similar way we can interpret the  profiles of the other authors.
For instance, authors C and D, who published the same number of papers (namely 40), have very similar profiles in terms of tendency to publish, but according to the proposed model, D has a higher tendency to collaborate, with a difference that also increases in time, and in particular the curve for D always dominates that for C; this is again coherent with the data in Table \ref{tab:authors}.
Finally, authors A and B have profiles which are in agreement with a smaller number of published papers that, at the same time, tends to increase from 2003 to 2012.

\begin{figure}[ht!]\centering
\includegraphics[width=8cm]{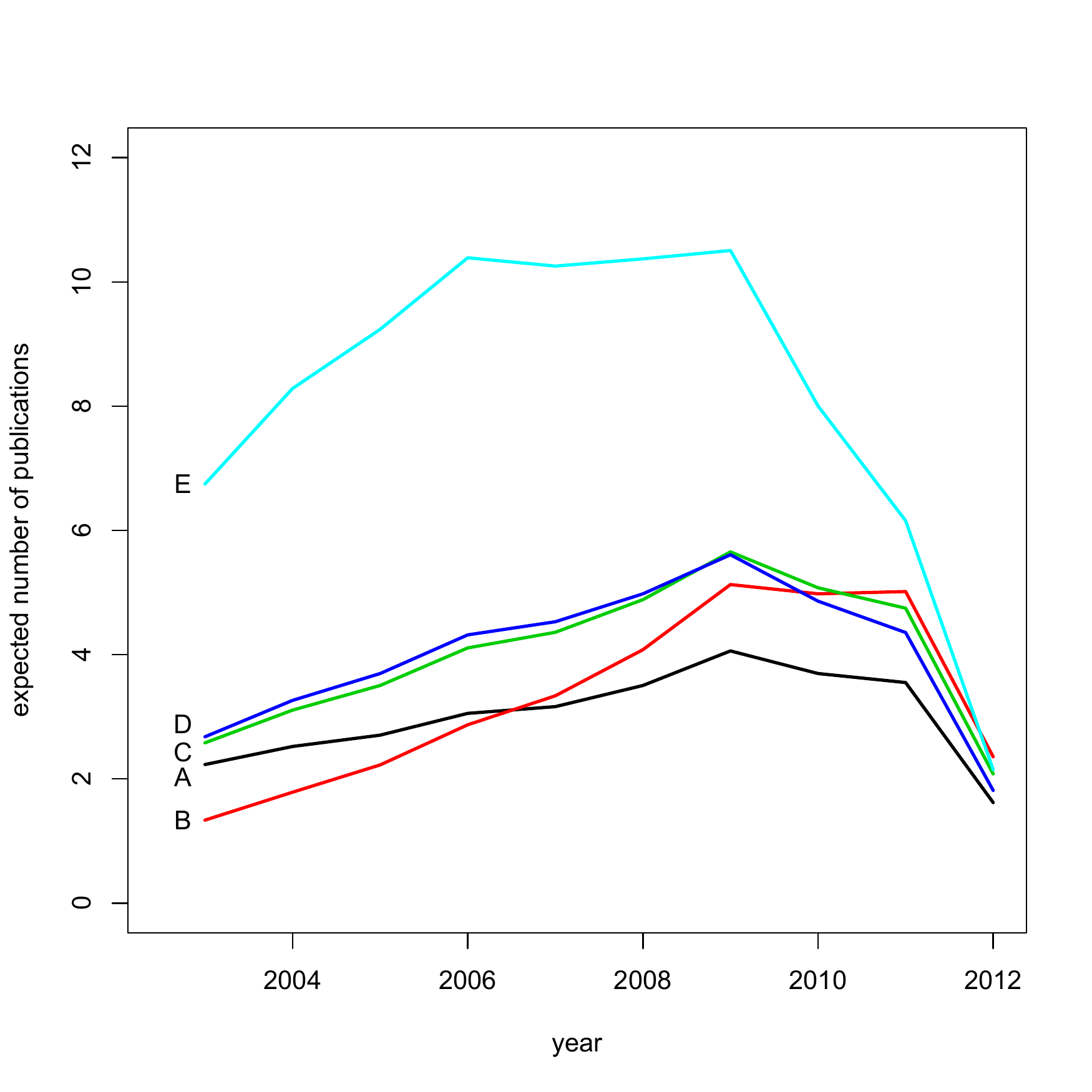}\caption{\em Profiles of tendency of authors to publish papers, measured by the expected number of yearly publications.}\label{fig:prof_ind2}
\end{figure}

To improve the interpretability of these profiles, instead of using the logits defined in (\ref{Eq:marginal1}), we can also express the tendency to publish in terms of expected number of publications per year. 
These expected values are obtained by multiplying the probability to be involved in a publication by the yearly number of publications available in Table \ref{tab:descriptive}.
The resulting profiles are reported in Figure \ref{fig:prof_ind2} and confirm the previous conclusions.

\begin{figure}[ht]\centering
\begin{tabular}{cc}
\includegraphics[width=8cm]{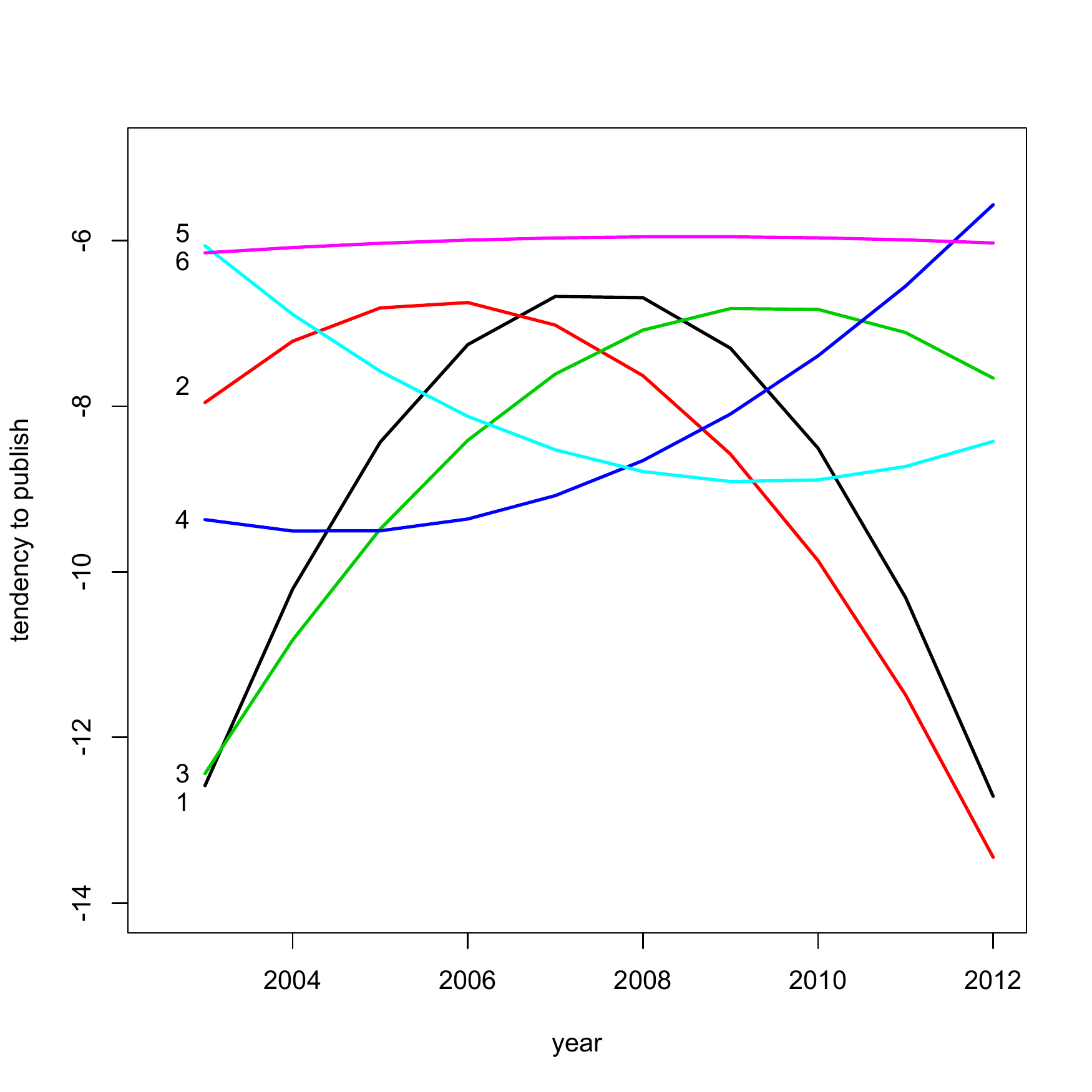} & \includegraphics[width=8cm]{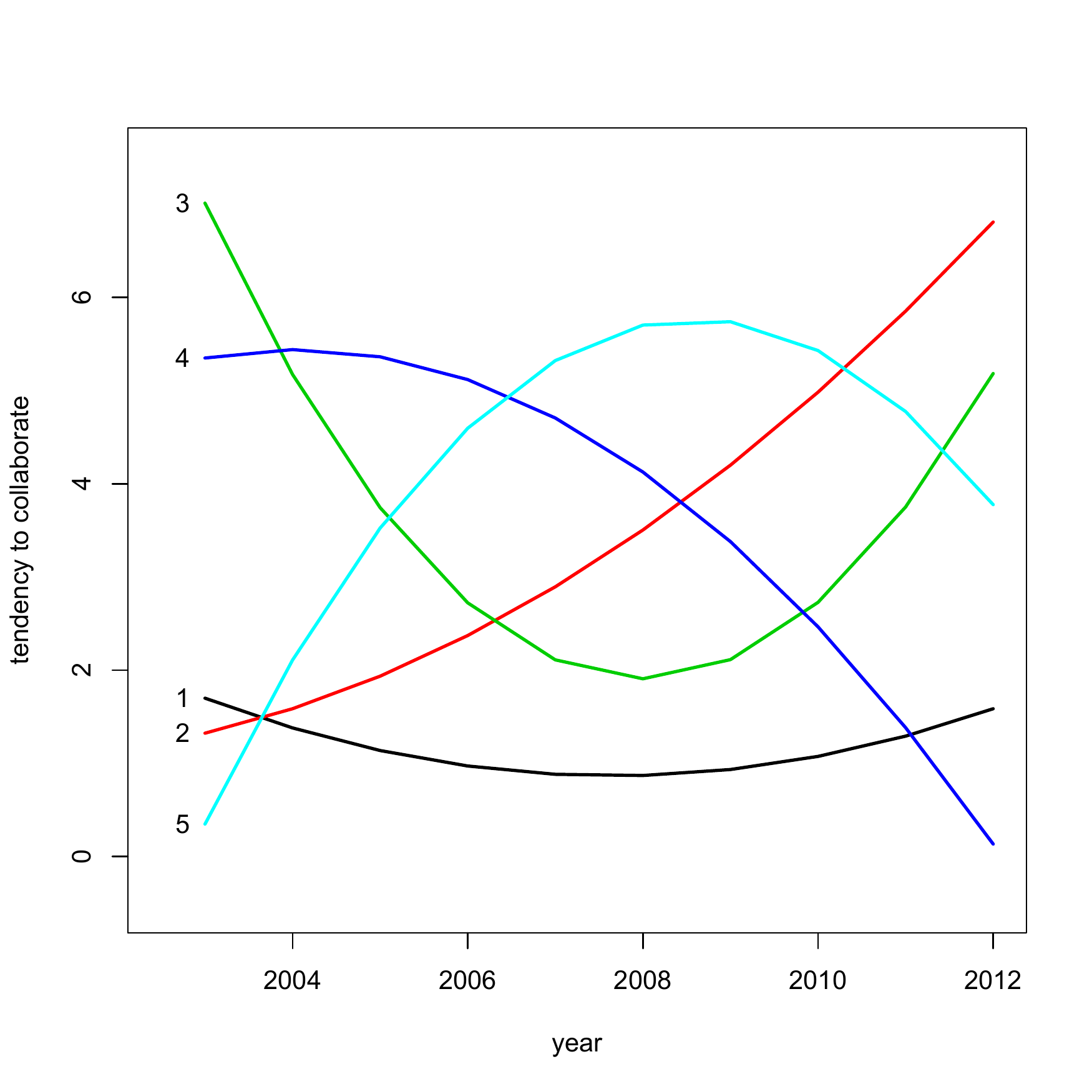}
\end{tabular}
\caption{\em Profiles in terms of tendency of authors to publish papers (left panel) and to collaborate (right panel), obtained by the cluster analysis}\label{fig:prof_clus}
\end{figure}

When we examine the overall sample of 3,607 authors, the analysis may be effectively carried out by building clusters of authors.
Following the approach described in Section \ref{sec:clustering}, we find evidence of $h_1=6$ different profiles in terms of tendency to publish and $h_2=5$ profiles in terms of tendency to collaborate.
The model with clustered profiles attains a maximum profile log-likelihood (normalized dividing by the number of ordered pairs of units) equal to -34.916 that is close that of the fixed-effects method, which is equal to -33.693.
On the other hand, the maximum pairwise log-likelihood with only one cluster is equal to -75.401.
This means that using a structure of $6\times 5$ clusters implies an improvement of the pairwise log-likelihood equal to 97.1\% with respect of using only one cluster, despite the huge reduction in the number of parameters with respect to the fixed-effects model: rather than using $n$ individual-specific parameter vectors $\b\al_i$ and $\b\be_i$, we use a very limited number of parameter vectors denoted by $\b\al^*_g$ and $\b\be^*_g$.
The corresponding profiles are represented in Figure \ref{fig:prof_clus}.

It is interesting to note that the 6 cluster profiles in terms of tendency to publish cover different possibilities.
In particular, profiles for the first 3 clusters have a reversed U-shape with picks located at different years.
On the other hand, profiles for clusters 4 and 5 have a U-shape but are rather different.
Finally, authors in cluster 6 tend to have a rather constant over time tendency to publish, which is higher than for the other clusters.
In any case, the values of the logit even for this class corresponds to low probability levels, with an expected number of publications which is smaller than 1 for all years.
In a similar way we can interpret the 5 clusters in terms of tendency to collaborate.
For instance, individuals in the first cluster have a general tendency to collaborate 
lower than the other authors, which is rather constant in time.

It is important to stress that, in principle, any author may belong to any cluster of the first type (in terms of tendency to publish) and of the second type (in terms of tendency to collaborate).
In order to better understand this aspect, we consider the cross classification of authors according to both criteria.
This cross classification is reported in Table \ref{tab:cross_class} that also shows the size of each cluster in terms of units assigned to it.

\begin{table}[ht]
\centering
\begin{tabular}{crrrrrr}
  \hline\hline
tend. to  & \multicolumn5c{tend. to collaborate}\\\cline{2-6}
publish & 1 & 2 & 3 & 4 & 5 & total\\ 
  \hline
  1 & 27 & 0 & 577 & 0 & 0 & 604 \\ 
  2 & 85 & 584 & 0 & 0 & 0 & 669 \\ 
  3 & 130 & 0 & 374 & 301 & 0 & 805 \\ 
  4 & 42 & 0 & 0 & 478 & 3 & 523 \\ 
  5 & 108 & 28 & 0 & 0 & 478 & 614 \\ 
  6 & 392 & 0 & 0 & 0 & 0 & 392 \\\hline
  total & 784 & 612 & 951 & 779 & 481 & 3607 \\    \hline
\end{tabular}
\caption{\em Cross classification of authors in terms of tendency to publish and to collaborate}\label{tab:cross_class}
\end{table}  

On the basis of the results in Table \ref{tab:cross_class} we observe that all clusters have a comparable dimension without a neat prevalence in terms of size of a specific cluster, although in terms of tendency to publish cluster 3 is the largest and the same happens for the tendency to collaborate.
Regarding the association between the two classification criteria, it is interesting to comment on certain regular patters that appear evident.
The most relevant one is that individuals in cluster 6 in terms of tendency to publish are all in the first cluster in terms of tendency to collaborate.
In summary, the authors with the highest tendency to publish have, at the same time, the smallest tendency to collaborate.
As noted above, when commenting the results shown in Figure \ref{fig:prof_ind} for author E, this means that, for the most productive authors, having a large number of coauthors is more plausibly ascribed to the tendency to publish than to a pure tendency to collaborate.
Author E can be considered as a pivotal or representative author for this joint class.

Other patterns may be easily discovered by looking at Table \ref{tab:cross_class} as, for instance, that authors in cluster 1 in terms of tendency to publish are mostly in cluster 3 in terms of tendency to collaborate.
Comparing the two profiles we observe that they have an opposite shape, which is coherent with the previous reasoning according to which the tendency to have a large number of coauthors may be reasonably ascribed to the general tendency to publish than to a specific social behavior.
This is likely due to the fact that scientific collaborations are viewed as long-term investments, and once a productive author establishes a team of researchers that effectively collaborate with each other, he/she tends to fully exploit these known scientific relations, with the aim of authoring even a large number of articles without changing the team.
\section{Conclusions}\label{sec:conclusions}
%
We propose a new model for 
social networks arising from a sequence of events involving an arbitrary (one or more) number of actors.
The main novelties, relative to available approaches, may be summarized as: ({\em i}) binary variables $Z_i^{(e)}$ for actor $i$ being involved in event $e$ are directly modeled instead of the tie variables $Y_{ij}^{(e)}=Z_i^{(e)}Z_j^{(e)}$; ({\em ii}) the model is based on marginal first- and second-order effects that have a meaningful interpretation in terms of tendency of an actor to participate in an event and tendency to cooperate; 
({\em iii}) these effects are parametrized accounting for each event time and individual fixed-effects parameters, so that the evolution of individual behaviors may be represented by suitable trajectories; ({\em iv}) inference is based on a composite likelihood function, built on the distribution of each ordered pair of units, and maximized with numerical complexity of order $O(n^2)$, $n$ being the network size; ({\em v}) units may be clustered in groups having the same behavior so as to simplify the interpretation of the data structure.

In conclusion, it is worth noting that the proposed approach may be potentially used in contexts different from our motivating example.
In particular, 
we can formulate 
third- 
(or higher) 
order effects to account for triangularizations.
This amounts to rely on a different composite likelihood function: a composite likelihood based on triples or a pairwise conditional likelihood, depending on the type of third-order effects used.
However, 
triples would necessarily 
lead to 
a higher computational complexity. 

Furthermore, 
individual covariates can be easily incorporated.
In fact, we rely on a parametrization based on a linear predictor 
with suitable polynomials of event times.
This linear predictor can 
include,
in a natural way, individual covariates
with no increase in complexity.
However, 
covariates specific to each pair of units (not to single units) and 
to events, 
contribute to
increase 
the computational burden.

Finally, 
our fixed-effect model
can be naturally extended to 
random-effects 
and lends itself to a Bayesian formulation.
By simply adding suitable priors on model parameters, the same inferential algorithm for finding the composite maximum likelihood estimate can be adopted to find maximum a posteriori estimates, with minor adjustments.

\section*{Acknowledgements}
We thank Daniele Durante for fruitful discussions. A. Mira and S. Peluso gratefully acknowledge financial support from Swiss National Science Foundation project 156229.

\section*{Appendix: Obtaining joint probabilities from marginal effects}
A simple but crucial issue concerns how to obtain the joint distribution of two binary variables starting from the corresponding marginal effects.
In particular, consider two binary variables $A$ and $B$ having joint probabilities
\[
p_{AB}(a,b):=p(A=a,B=b),\quad a,b=0,1,
\]
which are collected in lexicographical order in the vector $\b p$.
Also let $p_A(a):=p(A=a)$ and $p_B(b):=p(B=b)$ denote the corresponding marginal probabilities.
The marginal parameters are defined as
\begin{eqnarray*}
\eta_A &:=& \log\frac{p_A(1)}{p_A(0)},\\
\eta_B &:=& \log\frac{p_B(1)}{p_B(0)},\\
\eta_{AB} &:=& \log\frac{p_{AB}(0,0)p_{AB}(1,1)}{p_{AB}(0,1)p_{AB}(1,0)},
\end{eqnarray*}
and are collected, following the order given above, in the 3-dimensional vector $\b\eta$.

The inversion from $\b\eta$ to $\b p$ is based on a formula studied in a more general context by \cite{dale1986global}.
On the basis of some simple algebra, we find that the joint probability $p_{AB}(1,1)$ is equal to
\[
p_{AB}(1,1)=\frac{1+[p_A(1)+p_B(1)](e^{\eta_{AB}}-1)-\sqrt{\Delta}}{2(e^{\eta_{AB}}-1)},
\]
with
\[
\Delta = [1+(p_A(1)+p_B(1))(e^{\eta_{AB}}-1)]^2-4(e^{\eta_{AB}}-1)p_{A}(1)p_B(1)e^{\eta_{AB}}
\]
and where $p_A(1)=e^{\eta_A}/(1+e^{\eta_A})$, with $p_B(1)$ computed similarly.
Obviously, if $\eta_{AB}=0$, then the solution is simply $p_{AB}(1,1)=p_A(1)p_B(1)$.
In the end, we have
\[
\left\{
\begin{array}{l}
p_{AB}(1,0)=p_A(1)-p_{AB}(1,1),\\
p_{AB}(0,1)=p_B(1)-p_{AB}(1,1),\\
p_{AB}(0,0)=1-p_A(1)-p_B(1)+p_{AB}(1,1).
\end{array}
\right.
\]

Now consider the canonical (log-linear) parameters
\begin{eqnarray*}
\la_A &=& \log\frac{p_{AB}(1,0)}{p_{AB}(0,0)},\\
\la_B &=& \log\frac{p_{AB}(0,1)}{p_{AB}(0,0)},\\
\la_{AB} &=& \log\frac{p_{AB}(1,1)p_{AB}(0,0)}{p_{AB}(0,1)p_{AB}(1,0)},
\end{eqnarray*}
which are collected, following the order given  above, in the 3-dimensional vector $\b\la$, and note that
\[
p_{AB}(a,b) = \frac{e^{a\la_A+b\la_B+ab\la_{AB}}}{K(\b\la)},
\]
where $K(\b\la)$ is the normalizing constant.
For estimation purposes, the derivative of $\b\la$ with respect to $\b\eta$ is necessary.
We obtain this as the inverse of the derivative of $\b\eta$ with respect to $\b\la$, noting that
\begin{eqnarray*}
\eta_A &=& \la_A+\log\frac{1+e^{\la_B+\la_{AB}}}{1+e^{\la_B}},\\
\eta_B &=& \la_B+\log\frac{1+e^{\la_A+\la_{AB}}}{1+e^{\la_A}},\\
\eta_{AB} &=& \la_{AB}.
\end{eqnarray*}

Then we have:
\begin{equation}
\frac{\pa\b\eta}{\pa\b\la'}=
\left(\begin{matrix}
1 & \displaystyle{-\frac{e^{\la_B}(1-e^{\la_{AB}})}{(1+e^{\la_B})(1+e^{\la_B+\la_{AB}})}} & 
\displaystyle{-\frac{e^{\la_B+\la_{AB}}}{1+e^{\la_B+\la_{AB}}}} \\
\displaystyle{-\frac{e^{\la_A}(1-e^{\la_{AB}})}{(1+e^{\la_A})(1+e^{\la_A+\la_{AB}})}} & 1 &
\displaystyle{-\frac{e^{\la_A+\la_{AB}}}{1+e^{\la_A+\la_{AB}}}} \\
0 & 0 & 1
\end{matrix}\right),\label{eq:der}
\end{equation}
whereas regarding the inverse function we have
\[
\frac{\pa\b\la}{\pa\b\eta'}=\left(\frac{\pa\b\eta}{\pa\b\la'}\right)^{-1}=\left(1-\frac{\pa\eta_A}{\pa\la_B}\frac{\pa\eta_B}{\pa\la_A}\right)^{-1}
\left(\begin{matrix}
1 & \displaystyle{-\frac{\pa\eta_A}{\pa\la_B}} & 
\displaystyle{\frac{\pa\eta_A}{\pa\la_B}\frac{\pa\eta_{AB}}{\pa\la_B}-\frac{\pa\eta_{AB}}{\pa\la_A}} \\
\displaystyle{-\frac{\pa\eta_B}{\pa\la_A}} & 1 &
\displaystyle{\frac{\pa\eta_B}{\pa\la_A}\frac{\pa\eta_{AB}}{\pa\la_A}-\frac{\pa\eta_{AB}}{\pa\la_B}} \\
0 & 0 & \displaystyle{1-\frac{\pa\eta_A}{\pa\la_B}\frac{\pa\eta_B}{\pa\la_A}}
\end{matrix}\right),
\]
having elements directly taken from (\ref{eq:der}).


\end{document}